\documentclass[aps,superscriptaddress,pof,twocolumn,longbibliography]{revtex4-1}
\usepackage{graphicx}                                   % Include figure files
\usepackage{dcolumn}                                    % Align table columns on decimal point
\usepackage{bm}                                         % Bold math
\usepackage{bm}        % for math
\usepackage{amssymb}   % for math
\usepackage{babel}
\usepackage{verbatim}
\usepackage{amsmath}
\usepackage[export]{adjustbox}

\usepackage{siunitx}               % SI 
\begin{document}
	
	\title{Single-shot readout of a solid-state spin in a decoherence-free subspace}

\author{D. Farfurnik}
\affiliation{Department of Electrical and Computer Engineering, Institute for Research in Electronics and Applied Physics, and Joint Quantum Institute, University of Maryland, College Park, MD 20742, USA}

\author{R. M. Pettit}
\affiliation{Department of Electrical and Computer Engineering, Institute for Research in Electronics and Applied Physics, and Joint Quantum Institute, University of Maryland, College Park, MD 20742, USA}
\affiliation{Intelligence Community Postdoctoral Research Fellowship Program, University of Maryland, College Park, MD 20742}

\author{Z. Luo}
\affiliation{Department of Electrical and Computer Engineering, Institute for Research in Electronics and Applied Physics, and Joint Quantum Institute, University of Maryland, College Park, MD 20742, USA}

\author{E. Waks}

\affiliation{Department of Electrical and Computer Engineering, Institute for Research in Electronics and Applied Physics, and Joint Quantum Institute, University of Maryland, College Park, MD 20742, USA}

\begin{abstract}
The efficient single photon emission capabilities of quantum dot molecules position them as promising platforms for quantum information processing. Furthermore, quantum dot molecules feature a ``decoherence-free" subspace that enables spin qubits with long coherence time. To efficiently read out the spin state within this subspace requires optically cycling isolated transitions that originate from a triplet manifold within the quantum dot molecule. We propose and theoretically study a two-stage spin readout protocol within this decoherence-free subspace that allows single-shot readout performance. The process incorporates a microwave $\pi$-pulse and optically cycling the isolated transitions, which induces fluorescence that allows us to identify the initial spin state. This protocol offers enhanced readout fidelity compared to previous schemes that rely on the excitation of transitions that strongly decay to multiple ground states or require long initialization via slow, optically forbidden transitions. By simulating the performance of the protocol, we show that an optimal spin readout fidelity of over 97\% and single-shot readout performance are achievable for a photon collection efficiency of just 0.12\%. This high readout performance for such realistic photon collection conditions within the decoherence-free subspace expands the potential of quantum dot molecules as building blocks for quantum networks.
\end{abstract}
	\maketitle

Quantum dots emit single photons with high efficiency and indistinguishability \cite{Santori2001,Gazzano2013,Ding2016,Somaschi2016,Scholl2019}. By utilizing novel hybrid integration techniques to incorporate quantum dots on chip \cite{Elshaari2017,Davanco2017,Kim2017}, these optical properties position quantum dots as promising platforms for quantum computation \cite{Obrien2007} and communication \cite{Gisin2007}. Furthermore, electrically charged quantum dots offer a ground state spin qubit. Strongly coupling this qubit to photonic structures results in spin-selective reflectivity \cite{Carter2013,Sun2018,Luo2019,Najer2019}, which could pave the way toward the realization of quantum networks \cite{Kimble2008}. However, the potential to manipulate and store quantum information with quantum dot spin is limited due to the short coherence times that stem from interactions with the nuclear bath \cite{Bechtold2015,Stockill2016}. 

Quantum dot molecules, consisting of coupled quantum dots \cite{Stinaff2006,Kim2011,Elzerman2011}, offer a pathway to extend these coherence times by several orders of magnitude \cite{Weiss2012,Chow2016}. These molecules support singlet and triplet ground states that constitute a decoherence-free subspace, in which a spin qubit features reduced susceptibility to magnetic and electric field fluctuations. The conventional approach for reading the quantum dot molecule spin state within this subspace involves the resonant excitation of optical transitions and detection of the transmitted or reflected photons \cite {Kim2011}. However, the achievable readout fidelity utilizing such methods is limited as the excited states can decay to different ground states, preventing single-shot spin readout of the original spin state. An alternative spin readout approach involves repeatedly exciting an isolated transition of the quantum dot molecule that mainly decays to a single ground state \cite{Weiss2012,Delley2015}. As a result of such optical cycling, the number of photons collected within the measurement timeframe prior to the collapse to other ground states is significantly amplified, thereby increasing the spin readout fidelity. Accessing these isolated transitions from the decoherence-free subspace, however, has only been achieved by exciting weak optically forbidden transitions \cite{Weiss2012,Delley2015}. This approach features long spin transfer times that unnecessarily lengthen the readout process. Additionally, undesired decay through optically forbidden transitions during optical cycling of the isolated transitions diminishes the spin readout fidelity.

In this letter, we propose a protocol for reading the quantum dot molecule spin state within the decoherence-free subspace that does not rely on the optically forbidden transitions, enabling direct, efficient access to the isolated energy levels. The protocol involves two stages: First, a microwave $\pi$-pulse within the triplet manifold selectively transfers the spin state from the decoherence-free subspace to the ground levels of the isolated transitions. Second, the transitions are optically cycled and the emitted fluorescence is collected to identify the initial state. We simulate the spin readout fidelities achievable through the implementation of this protocol. Additionally, we identify the readout times and photon collection efficiencies that provide single-shot spin readout capabilities. Spin readout fidelities of over 97\% and single-shot readout performance simulated for photon collection efficiencies as low as 0.12\% are expected to be feasible by delivering the microwave pulse utilizing a coplanar waveguide. Such spin readout performance strengthens the potential of quantum dot molecules for quantum information processing.

\begin{figure}[h]
	\centering
	\includegraphics[width=0.235\textwidth]{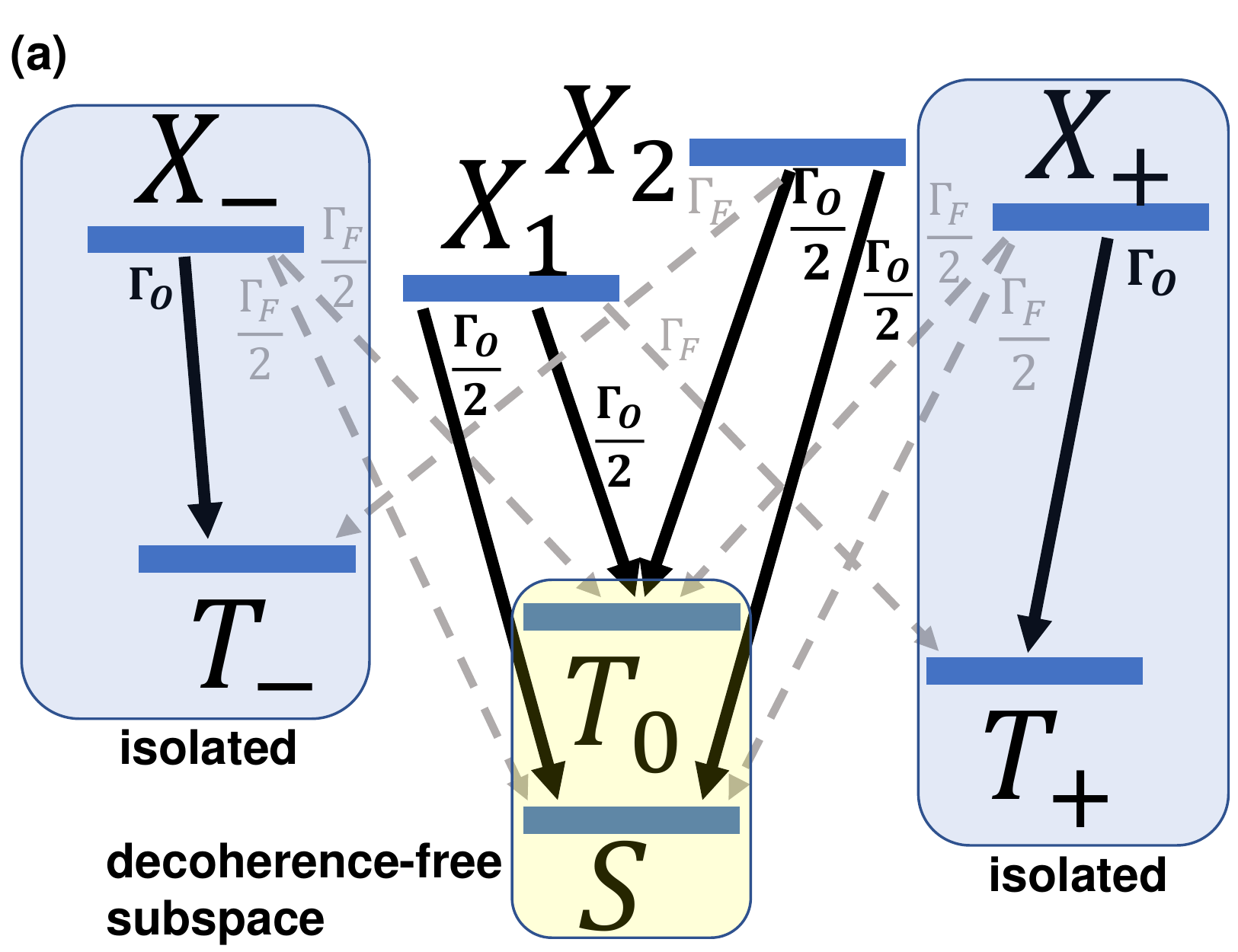}
	\includegraphics[width=0.235\textwidth]{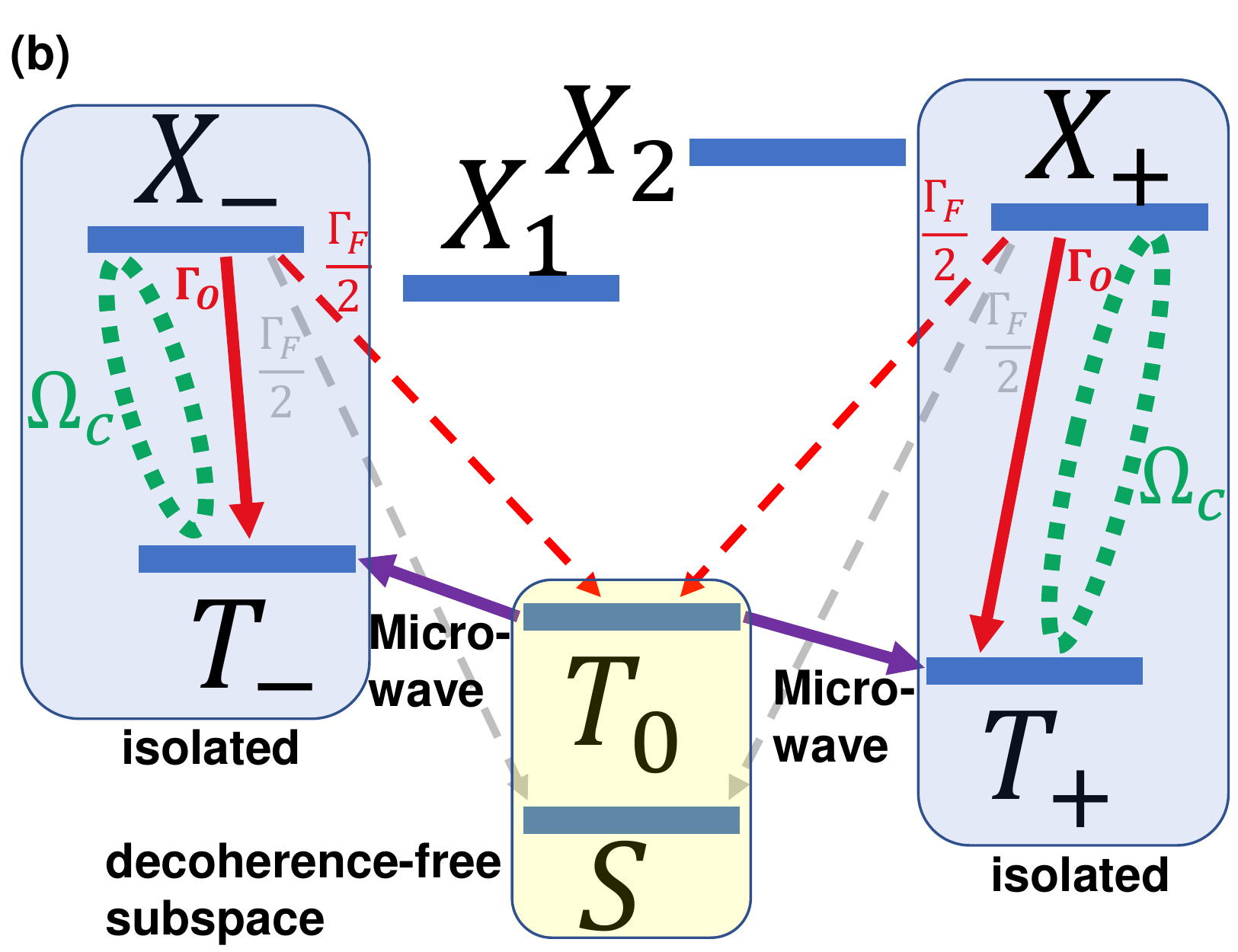}
	\caption{(Color online) (a) Energy-level diagram of a quantum dot molecule. The solid black arrows represent optically allowed transitions; the dashed grey arrows represent optically forbidden transitions. (b) Same diagram, showing only the transitions relevant for the spin readout protocol. First, the $T_0$ state is transferred to either $T_+$ or $T_-$ (purple arrows) by a microwave $\pi$-pulse. Then, the isolated transitions $T_+\leftrightarrow X_+$ and $T_-\leftrightarrow X_-$ are optically cycled (green dotted lines) and the fluorescence emitted to the triplet manifold (red arrows) is collected. Undesired decay to the $S$ ground state via optically forbidden transitions (grey arrows) eventually limits the readout performance. }
	\label{fig:fig1}
\end{figure}

Our protocol considers a singlet-triplet energy level structure associated with quantum dot molecules that contain a single electron in each quantum dot. The application of magnetic field parallel to the sample growth direction (Faraday geometry) Zeeman-splits the quantum dot molecule energy levels to form singlet ($\lvert S\rangle=\lvert \uparrow\rangle\lvert\downarrow\rangle-\lvert \downarrow\rangle\lvert\uparrow\rangle$) and triplet ($\{\lvert T_0\rangle=\lvert\uparrow\rangle\lvert\downarrow\rangle+\lvert \downarrow\rangle\lvert\uparrow\rangle,\lvert T_+\rangle=\lvert\uparrow\rangle\lvert\uparrow\rangle,\lvert T_-\rangle=\lvert \downarrow\rangle\lvert\downarrow\rangle\} $) spin manifolds, as well as four optically excited states ($\{X_-,X_+,X_1,X_2\}$)  [Fig. \ref{fig:fig1} (a)] \cite{Stinaff2006,Kim2011,Elzerman2011}. The spin manifold $\{S,T_0\}$ has a reduced susceptibility to fluctuations in the nuclear bath due to the lack of electric and magnetic dipole moments, thereby forming a decoherence-free subspace \cite{Kim2011,Elzerman2011}. Optical selection rules allow fast spontaneous emission at a rate of $\Gamma_O$ through spin conserving optical transitions [black solid lines in Fig. \ref{fig:fig1} (a)], while optically forbidden transitions, weakly enabled by heavy-light hole mixing [grey dashed lines in Fig. \ref{fig:fig1} (a)], exhibit a significantly slower decay rate of $\Gamma_F$ \cite{Weiss2012,Delley2015}.

The spin state within the decoherence-free subspace can be read by exciting any of the four optically allowed transitions leading from the subspace to the excited states $X_1$ and $X_2$, and collecting the resulting transmitted or reflected resonant fluorescence \cite {Kim2011}. However, since the excited states decay to both spin states $S$ and $T_0$, both spin states will become equally populated before enough photons are collected for the identification of the original state, thereby preventing single-shot spin readout. An alternative approach for spin readout involves the excitation of the optical transitions $T_+\leftrightarrow X_+$ and $T_-\leftrightarrow X_-$, which are isolated from other energy levels \cite{Weiss2012,Delley2015}. Due to the isolation of these transitions, the number of photons detected before the state decays to the singlet state is significantly amplified, thereby enabling single-shot spin readout. However, leveraging these transitions requires a preliminary stage of spin transfer from the decoherence-free subspace to $T_+$ or $T_-$. To date, such transfer has been demonstrated utilizing forbidden transitions leading to the excited states $X_-$  and $X_+$. However, the slow rates of these transitions result in long ($\sim$ microseconds) spin transfer times, lengthening the spin readout process without contributing to the performance. Furthermore, residual decay via these forbidden transitions to the $S$ state during optical cycling eventually limits the number of photons collected via the isolated transitions, thereby reducing the overall performance of the readout protocol.

We propose a two-stage protocol for improving the spin state readout of a qubit within the decoherence-free-subspace of a quantum dot molecule that does not depend on forbidden transitions [Fig. \ref{fig:fig1} (b)]. First, a microwave $\pi$-pulse transfers the $T_0$ state from the decoherence-free subspace to one of the complementary triplet states, $T_+$ or $T_-$ [purple arrows in Fig. \ref{fig:fig1} (b)]. As the transitions $T_0\rightarrow T_+$ and $T_0\rightarrow T_-$ are degenerate, this pulse will lead to the occupation of either of the states, $T_+$  or $T_-$, with equal probabilities. Then, two simultaneous laser pulses (i.e., the “readout pulses," with a duration of $T_R$) cycle both isolated transitions of $T_+\leftrightarrow X_+$ and $T_-\leftrightarrow X_-$ [green dotted lines in Fig. \ref{fig:fig1} (b)] and the emitted fluorescence decaying to the triplet manifold is collected [red arrows in Fig. \ref{fig:fig1} (b)]. The simultaneous driving of both transitions ensures optical excitation regardless of the specific previously occupied state ($T_+$  or $T_-$). Due to the isolation of these transitions, the resulting emitted fluorescence is significantly amplified compared to protocols relying on excited states strongly decaying to both triplet and singlet manifolds. Collecting this fluorescence signal, which should be large for the initial state $T_0$ and small for the initial state $S$, will indicate the original spin state within the decoherence-free subspace. To avoid the need for complex spectral filtering, we consider the collection of a broadband signal incorporating emissions from all transitions decaying to the triplet manifold (including the forbidden ones) [red arrows in Fig. \ref{fig:fig1} (b)]. 

\begin{figure}[h]
	\centering
	\includegraphics[width=0.235\textwidth]{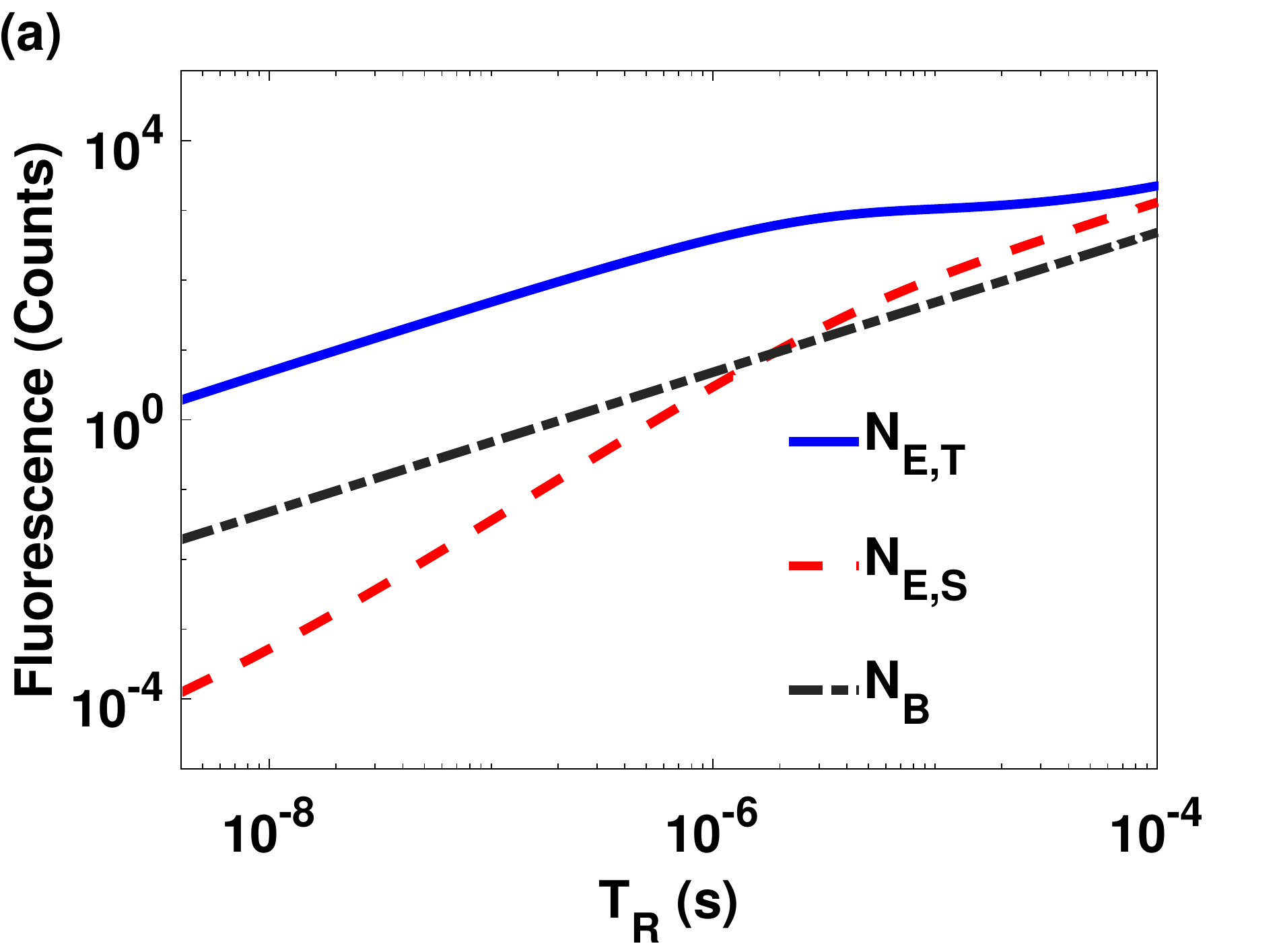}
	\includegraphics[width=0.235\textwidth]{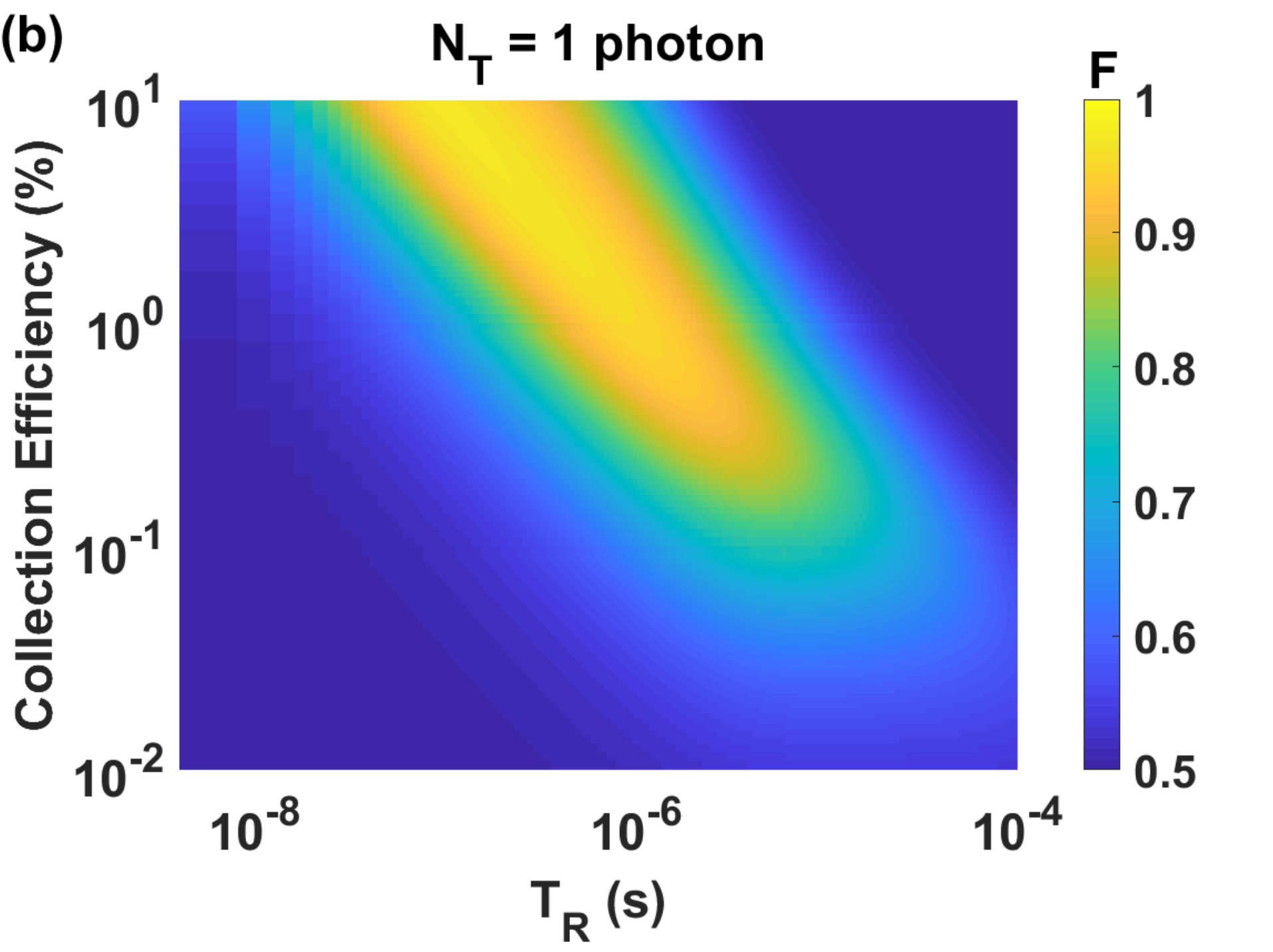}
	\includegraphics[width=0.235\textwidth]{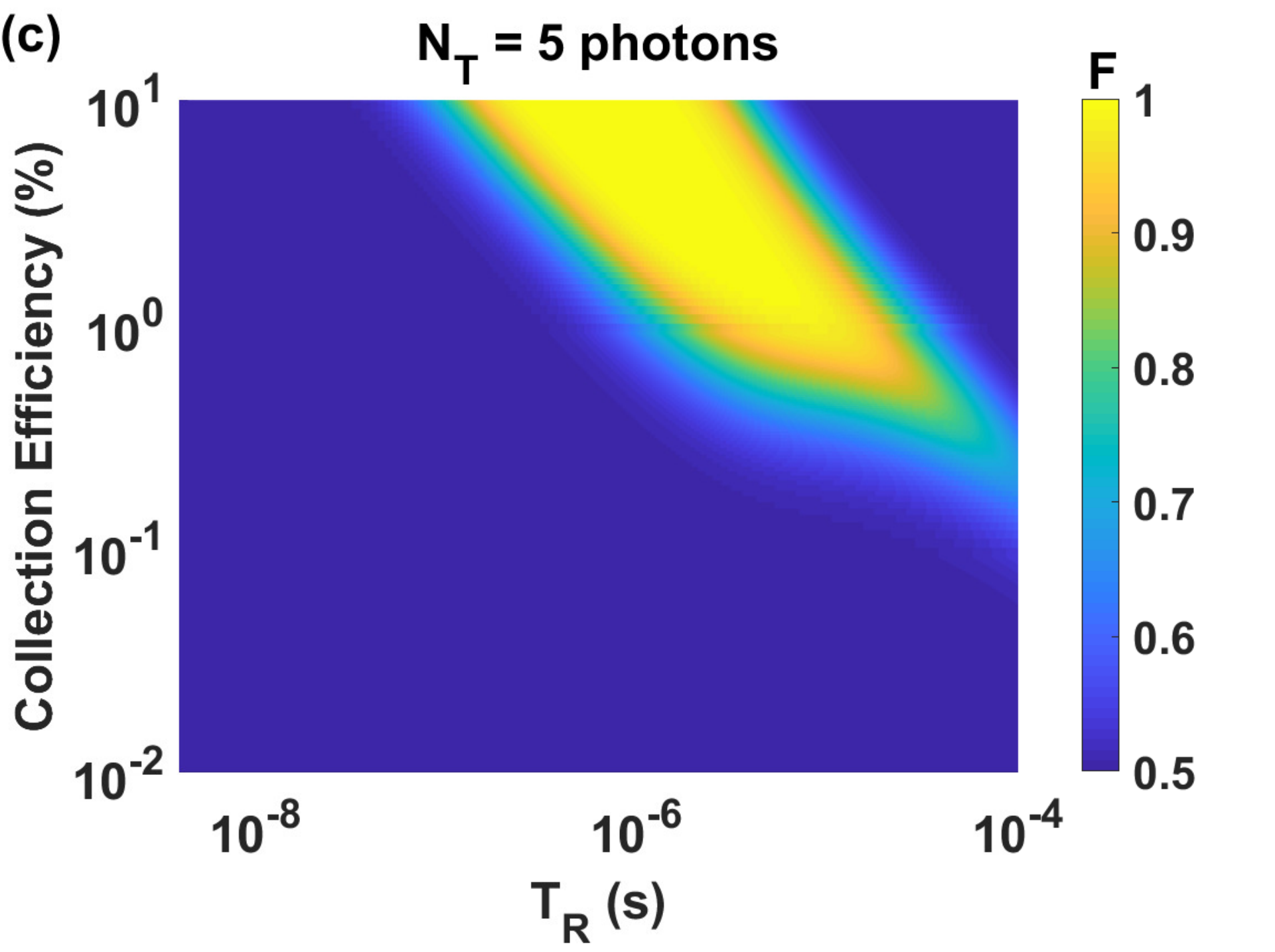}
	\includegraphics[width=0.235\textwidth]{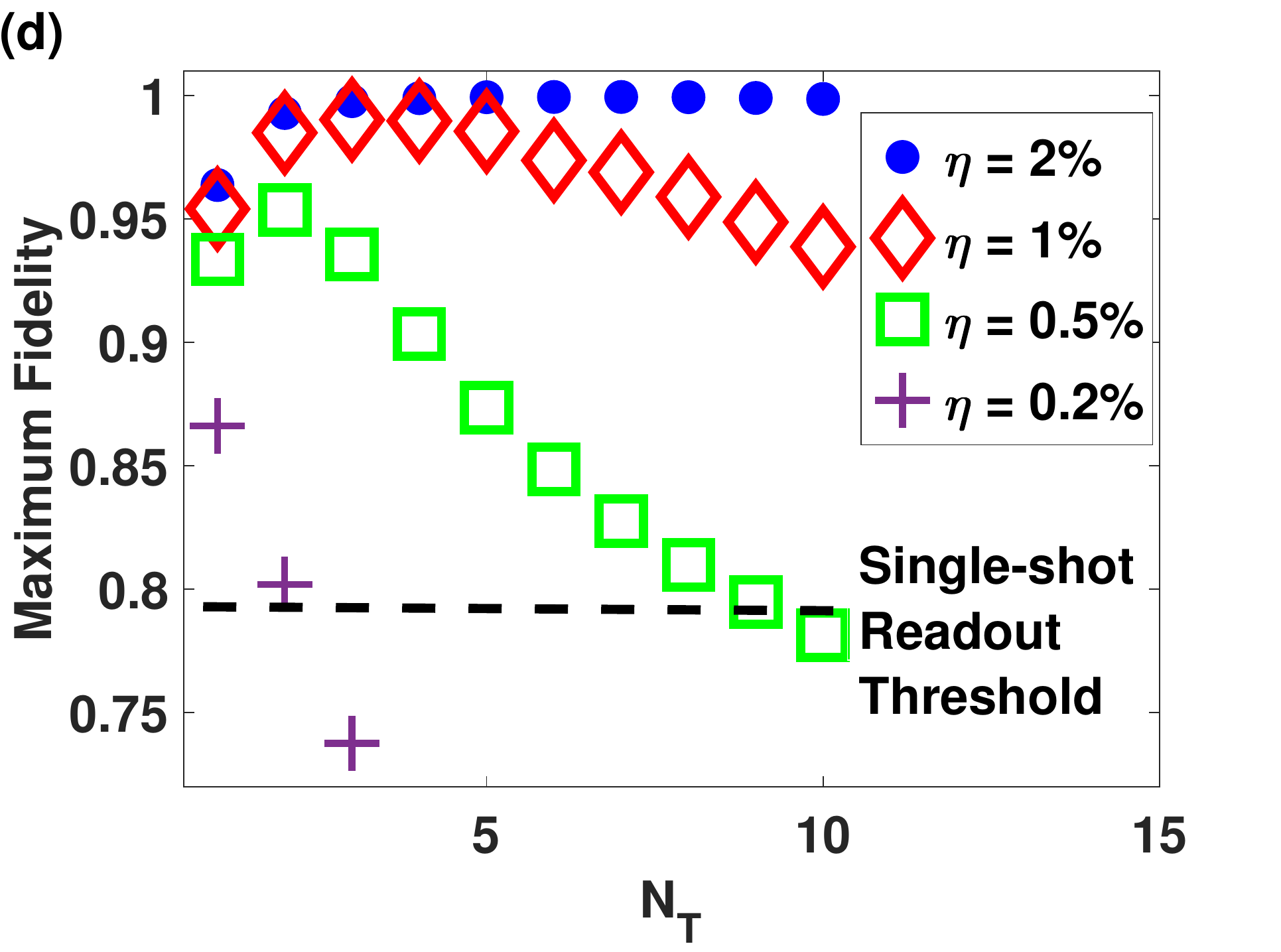}
	
	\caption{(Color online) (a) Simulated fluorescence signal emitted from a quantum dot molecule ($\Gamma_F=1$ MHz) for the initial $T_0$ ($N_{E,T}$, blue solid line) and $S$ ($N_{E,S}$, red dashed line) states, as well as the signal reflected from the sample considering a cross polarization degree of $10^6$ ($N_B$, black dash-dotted line) as a function of the duration of the readout pulses. (b)-(c) Simulated spin readout fidelity (F) as a function of the collection efficiency and duration of the readout pulses ($\Gamma_F=1$ MHz) at (b) 1-photon and (c) 5-photon collection thresholds ($N_T$). (d) Maximum spin readout fidelity as a function of the collection threshold for different collection efficiencies. The single-shot readout threshold is indicated by the black dashed line. }
	\label{fig:fig2}
\end{figure}

To explore the protocol's ability to identify the initial spin state, we first simulate the emitted and reflected signals as a function of the duration ($T_R$) of the readout pulses [Fig. \ref{fig:fig2} (a)] by solving the Lindblad master equation for the density matrix $\rho$,  
\begin{equation} \label{eq:1}
\dot{\rho}=-\frac{i}{\hbar}\left[H,\rho\right]+\sum_i  \Gamma_i\left( L_i \rho L_i^\dagger-\frac{1}{2} \left\{L_i^\dagger L_i,\rho \right\}  \right),
\end{equation} 
where $H$ is the Hamiltonian incorporating the microwave $\pi$-pulse, and $\{L_i\}$ are jump operators of incoherent processes with decay rates of $\{\Gamma_i\}$. These processes include optically allowed ($\Gamma_O=1$ GHz) and forbidden (100 kHz - 10 MHz) transition decay rates, ground state spin depolarization ($\frac{1}{\Gamma_1} =100$  $\SI{}{\micro\second}$), spin decoherence within the singlet-triplet qubit ($\frac{1}{\Gamma_{2,ST}} =200$ ns) and triplet manifold ($\frac{1}{\Gamma_{2,T}} =5$ ns), and optical cycling associated with the protocol ($\Omega_C=1$ GHz) (full description in \cite{Suppl}). Moreover, the simulations consider direct reflectivity of a 1 $\SI{}{\micro\watt}$ laser from the sample surface attenuated by a factor of $10^6$, which is achievable using cross-polarized optics. The numerical values were based on measurements of indium arsenide quantum dots embedded in gallium arsenide \cite{Weiss2012,Delley2015,Luo2019,Lu2010,Press2008,Bodey2019,Suppl}.

The simulation results of Eq. \eqref{eq:1} [Fig. \ref{fig:fig2} (a)] highlight the difference between the number of photons emitted for the initial spin states $\lvert T_0 \rangle$ ($N_{E,T}$) and $\lvert S \rangle$ ($N_{E,S}$). For the initial $\lvert T_0\rangle$ state (blue solid line), the fluorescence intensity mainly depends on the branching ratio between the optically allowed and forbidden transitions, $\frac{\Gamma_O}{\Gamma_F}$. Ideally, as the $\lvert S\rangle$ state should not be transferred to the isolated manifold by the preliminary microwave $\pi$-pulse, zero fluorescence should be expected for this initial state. However, an undesired signal (red dashed line) emerges due to spin depolarization to the triplet manifold at non-zero temperatures, thereby reducing the spin readout fidelity. Additionally, signal directly reflected from the sample ($N_B$) (black dash-dotted line) contributes to the background photon count for both initial spin states, which also reduces the spin readout fidelity. A significant drop in the ability to distinguish between the spin states is observed at increasing durations. This drop can be attributed to the fluorescence signals of both initial states approaching each other [red and blue lines in Fig. \ref{fig:fig2} (a)] due to spin depolarization and to the increase of the direct reflectivity [black dash-dotted line in Fig. \ref{fig:fig2} (a)].   

We define a photon collection threshold value (an integer, $N_T$) to determine the spin readout fidelity within a single realization of the protocol. If the number of photons collected is greater or smaller than $N_T$, then we classify the initial spin state as $\lvert T_0\rangle$ or $\lvert S\rangle$, respectively \cite{Danjou2014}. For an initial state within the $\{S,T_0\}$ manifold, the resulting experimentally achievable readout fidelity, which strongly depends on the optical collection efficiency, is defined by \cite{Danjou2014,Delteil2014}
\begin{equation} \label{eq:2}
F=1-\frac{1}{2}\left(P_{T,F}+P_{S,F}\right),
\end{equation} 
where $P_{T,F}$ ($P_{S,F}$) is the false negative (positive) probability of detecting less than (at least) $N_T$ photons given the initial state $\lvert T_0\rangle$ ($\lvert S\rangle$). These probabilities can be calculated from the Poissonian distribution of the photon detection process,
\begin{equation} \label{eq:3}
P_{T,F}=\sum_{k=1}^{N_T-1} \frac{\lambda_T^k e^{-\lambda_T}}{k!} \quad ; \quad
P_{S,F}=\sum_{k=N_T}^{\infty} \frac{\lambda_S^k e^{-\lambda_S}}{k!}
\end{equation} 
with expected values of $\lambda_j=\eta(N_{E,j}+N_B)$, where $\eta$ is the photon collection efficiency dictated by the experimental setup and $j\in\{S,T\}$. 

Next, we simulate the readout fidelity under the realization of the protocol [Eqs. \eqref{eq:2}-\eqref{eq:3}] as a function of the optical collection efficiency and duration of readout pulses for different photon thresholds, including $N_T=1$  [Fig. \ref{fig:fig2} (b)] and $N_T = 5$ [Fig. \ref{fig:fig2} (c)]. For a given photon threshold, certain parameter regimes [yellow regions in Fig. \ref{fig:fig2} (b)-(c)] result in low false negative probabilities of $P_{T,F}$ and $P_{S,F}$, corresponding to high readout fidelities. Parameter regimes exhibiting low fidelities [blue region in Fig. \ref{fig:fig2} (b)-(c)] arise either due to insufficient detection of photons (bottom-left region), or when many photons are collected for the initial $S$ state as a result of spin depolarization and direct reflectivity (top-right region). While defining a photon threshold of $N_T = 1$ provides high fidelities ($> 0.9$) at a broad range of durations and collection efficiencies [Fig. \ref{fig:fig2} (b)], defining a higher threshold, such as $N_T=5$, results in even greater achievable fidelities [Fig. \ref{fig:fig2} (c)]. Fig. \ref{fig:fig2} (d) illustrates the fidelities possible for different collection efficiencies. For example, $\eta=1\%$, which has been realized experimentally by incorporating quantum dot molecules in a low-Q cavity \cite{Delteil2014}, provides maximum readout fidelities of $F\sim 0.95$ for $N_T=1$ and $F\sim 0.99$ for $N_T=4$.

\begin{figure}[h]
	\centering
	\includegraphics[width=0.235\textwidth]{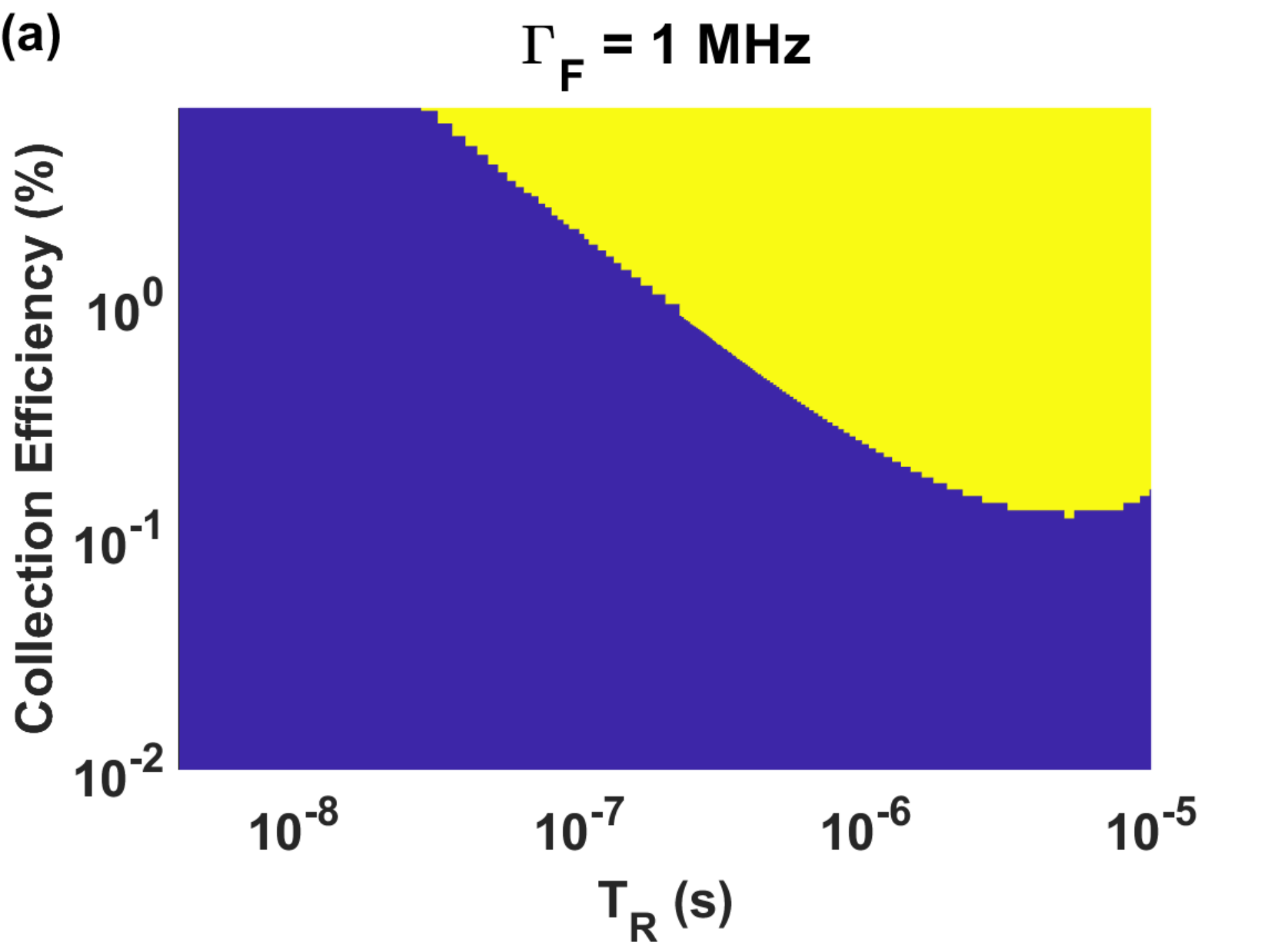}
	\includegraphics[width=0.235\textwidth]{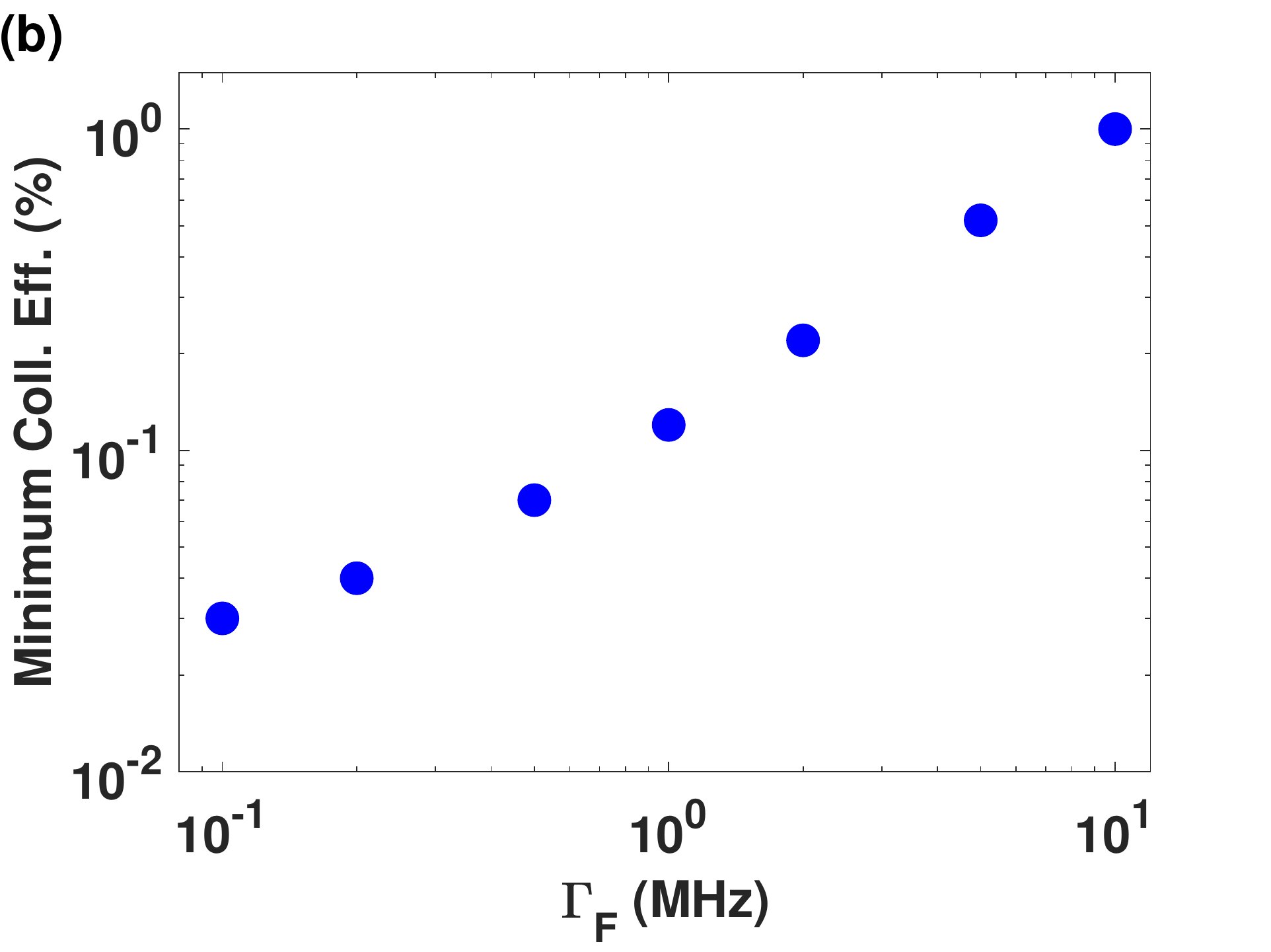}
	\caption{(Color online) (a) Simulated readout pulse durations and photon collection efficiencies required for single-shot readout ($\Gamma_F=1$  MHz). The yellow region indicates sets of parameters that result in a readout fidelity of $ F >0.8$. (b) The minimum collection efficiency for single-shot readout as a function of the forbidden transition decay rate. }
	\label{fig:fig3}
\end{figure}

A spin readout protocol is conventionally considered to offer single-shot capabilities when the readout fidelity exceeds 0.8 [Fig. \ref{fig:fig2} (d)] \cite{Danjou2014}. Under such conditions, the spin state detected within a single realization ($T_0$ or $S$) can be further used for quantum information processing. In Fig. \ref{fig:fig3} (a), we indicate in yellow the collection efficiency and readout duration regimes that could provide single-shot capabilities for a forbidden transition decay rate of $\Gamma_F=1$ MHz. Similarly, we extract the minimum collection efficiency required for single-shot readout considering different forbidden transition decay rates [Fig. \ref{fig:fig3} (b)] to account for sample variations based on the strength of heavy-light hole mixing. The parameter span that enables single-shot readout reduces significantly with increasing forbidden transition rates due to faster decay from the isolated manifold to the singlet state. However, for a previously measured rate of $\Gamma_F=5 $ MHz \cite{Delley2015}, single-shot readout is possible for collection efficiencies greater than 0.52\%, while similar capabilities for the weaker rate of $\Gamma_F=1$ MHz require collection efficiencies of just 0.12\%. Thus, the proposed protocol enables single-shot readout within a broad range of conditions. In particular, if no photons are detected within a readout realization, the initial spin state is identified as $S$ state with probability over $80\%$.     

Experimentally observing these high-performance spin readout capabilities requires the implementation of high-fidelity microwave $\pi$-pulses on a timescale approaching the decoherence time within the triplet manifold (representatively chosen as 5 ns \cite{Suppl}). Previous implementations of microwave control \cite{Kroner2008} could be improved utilizing coplanar waveguides, which enabled gigahertz Rabi frequencies for the electron spin of nitrogen-vacancy centers in diamond with a g-factor of 2 \cite{Fuchs2009}. Considering the quantum dot molecule spin g-factor of $\sim 0.5$ \cite{Weiss2012,Delley2015}, and the proportionality of the Rabi frequency to $I/r$, where $I$ is the current in the waveguide and $r$ is the distance to the spin (Ampere's law), analogous rotations of the quantum dot molecule spin should be achievable by doubling the microwave amplification and reducing the waveguide diameter.

\begin{figure}[h]
	\centering
	\includegraphics[width=0.235\textwidth]{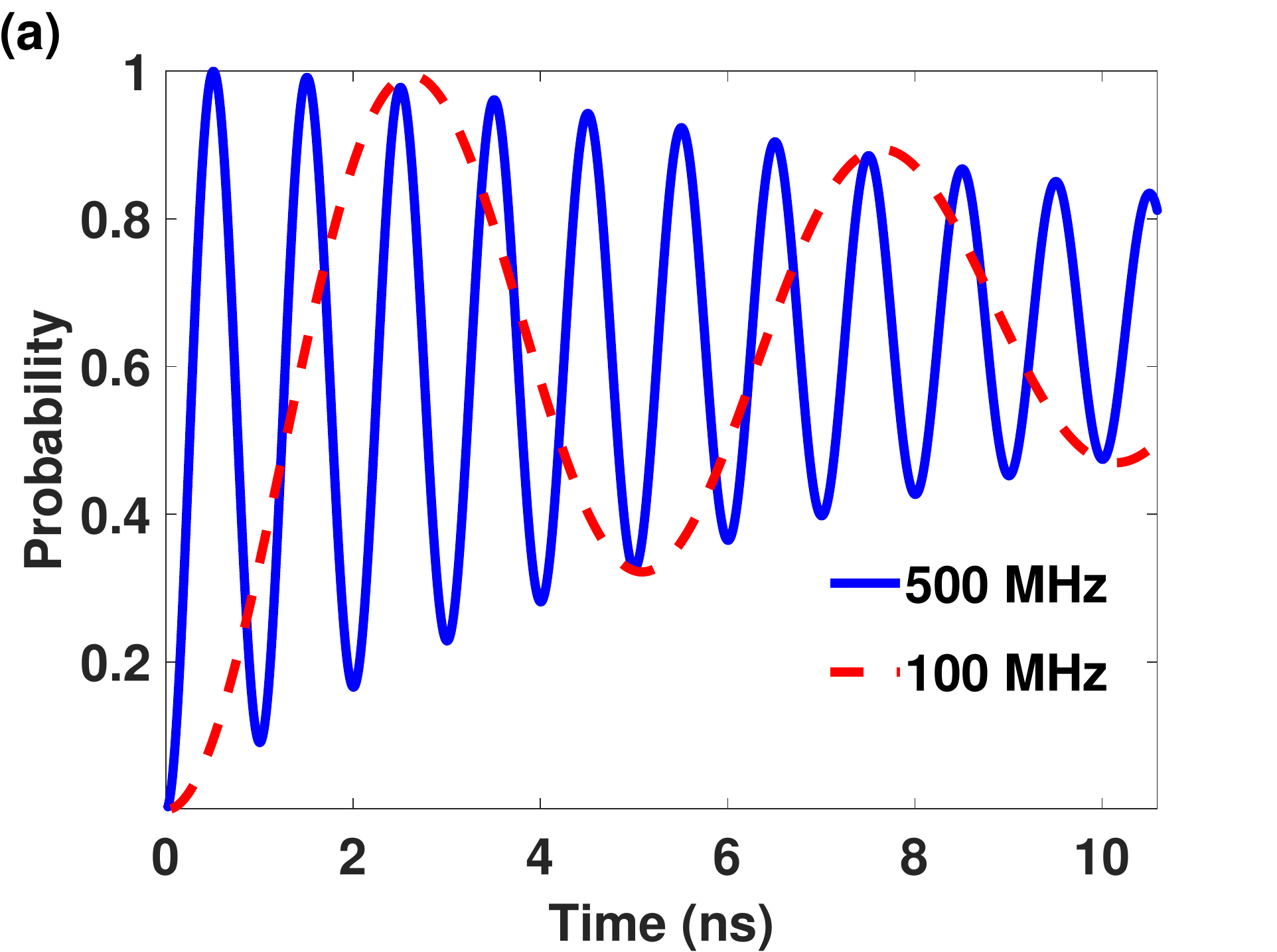}
	\includegraphics[width=0.235\textwidth]{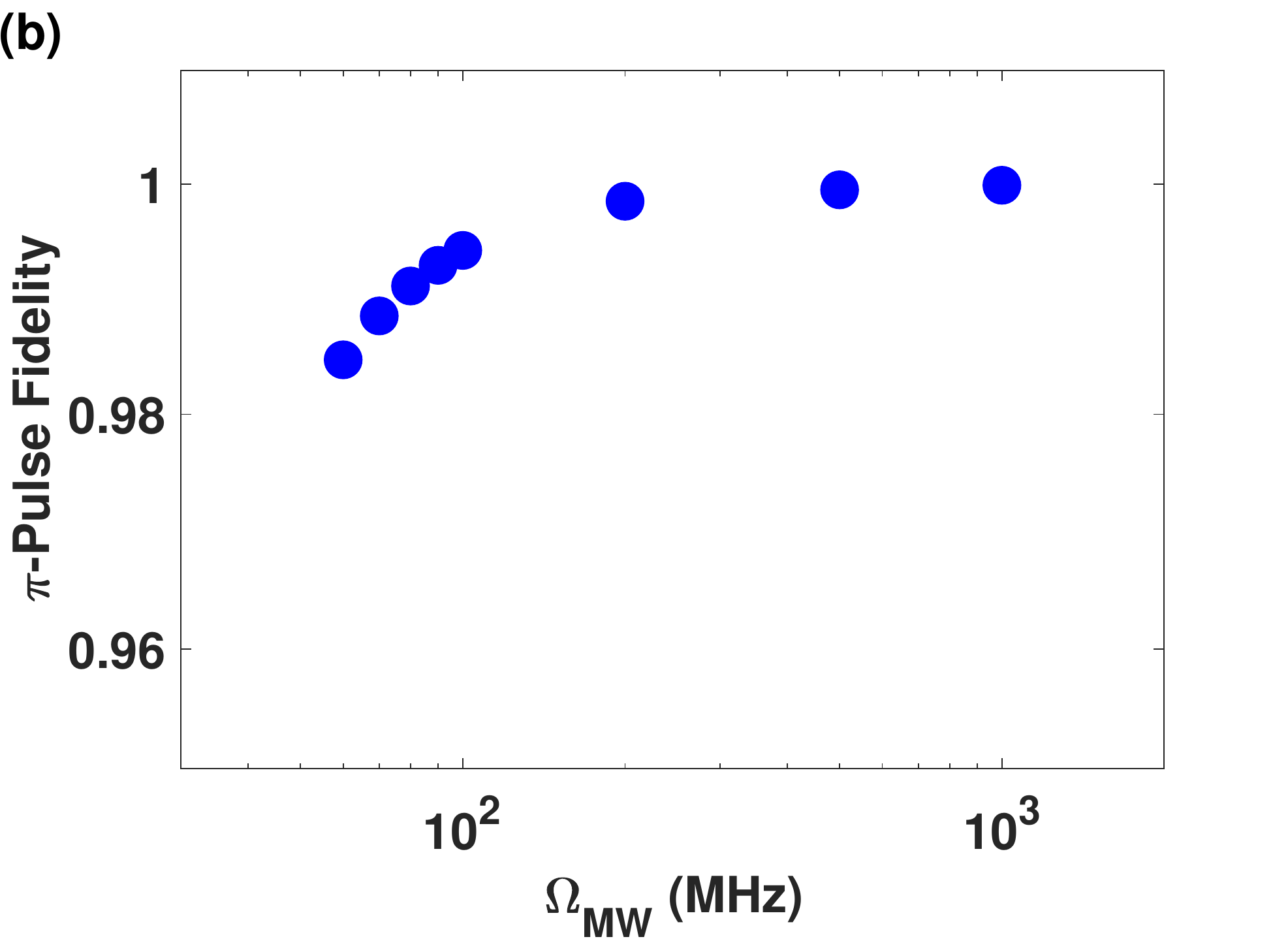}
	\caption{(Color online) (a) Simulated probability of populating the $T_+$ or  $T_-$ state while resonantly driving the initial $T_0$ state at different microwave Rabi frequencies. (b) The fidelity of the microwave $\pi$-pulse as a function of the Rabi frequency.}
	\label{fig:fig4}
\end{figure}

While the simulated fidelities depicted in Figures \ref{fig:fig2}-\ref{fig:fig3} consider ideal, infinitely strong microwave $\pi$-pulses, the experimental readout performance will degrade due to the finite Rabi frequency of realistic pulses, dictated by losses and impedance mismatch along the coplanar waveguide \cite{Suppl}. To quantify the resulting degradation of the readout fidelity, we simulate the probability of populating the $T_+$ or $T_-$ state after driving the initial state $T_0$ [Fig. \ref{fig:fig4} (a)] at different Rabi frequencies, $\Omega_{MW}$. For each Rabi frequency, the first maximum of the oscillating probability dynamics corresponds to the $\pi$-pulse fidelity, which can be multiplied by the fidelities provided in Fig. \ref{fig:fig2} to extract the total spin readout fidelity. Evidently, even Rabi driving at a frequency of 100 MHz [Fig. \ref{fig:fig4} (b)], almost an order of magnitude slower than achieved previously \cite{Fuchs2009}, results in a multiplication factor of $\sim 0.98$ and a total spin readout fidelity over 97\%. Thus, the experimental realization of such spin rotations could enable single-shot readout of the quantum dot molecule spin.

Additional factors that may limit the spin readout fidelity include finite experimental temporal resolution in applying the microwave and optical pulses, amplitude fluctuations of the microwave field \cite{Farfurnik2017cont}, and heating-induced depolarization caused by the applied microwave and laser fields \cite{Bodey2019}. While the effect of temporal jitter should be negligible in typical $\sim 20$ ps resolution time-resolved measurements, and microwave fluctuations can be suppressed by utilizing phase-modulated driving \cite{Farfurnik2017cont}, the non-trivial effect of heating strongly depends on the cryostation cooling power, which is subject to experimental studies beyond the scope of this work.

To summarize, we propose a microwave-based protocol for reading the state of a spin qubit within the decoherence-free-subspace of a quantum dot molecule. The readout fidelities presented here could be further improved by operating at sub-Kelvin temperatures, for which thermal equilibrium mostly populates the singlet state and spin depolarization is minimized. While the optical signal collected from implementing the protocol represents spin population (i.e., the expectation value $\langle S_z \rangle$), the full state tomography of the quantum state (i.e., expectation values $\langle S_x\rangle ,\langle S_y\rangle$) can be derived by utilizing a preceding optical Raman rotation $\frac{\pi}{2}$-pulse, as previously implemented on quantum dot systems \cite{Press2008,Bodey2019,Kim2011}. Combined with the long coherence times associated with the decoherence-free subspace, the presented single-shot readout capabilities for modest collection efficiencies could enhance the performance of quantum dots for generating photon entanglement and storing quantum information, thereby upgrading their potential as building blocks of quantum networks. Finally, the protocol can be implemented even if the transitions $T_0\leftrightarrow T_+$ and $T_0\leftrightarrow T_-$ are not degenerate (e.g. in the presence of zero-field splitting). In such a case, while photons will be collected mainly from exciting a single isolated manifold, the equivalence between both isolated manifolds ensures that the readout performance will not change compared to the one presented here. 
\section*{Acknowledgements}
 This work has been supported by the Physics Frontier Center at the Joint Quantum Institute, the National Science Foundation (Grants PHY1415485 and ECCS1508897), and the ARL Center for Distributed Quantum Information (Grant W911NF1520067). DF acknowledges support by the Fulbright Postdoctoral Fellowship and the Israel Council for Higher Education Quantum Science and Technology Scholarship. RMP acknowledges support through an appointment to the Intelligence Community Postdoctoral Research Fellowship Program at the University of Maryland, administered by Oak Ridge Institute for Science and Education through an interagency agreement between the U.S. Department of Energy and the Office of the Director of National Intelligence.

\bibliography{../../mytex/mybibliography}
\end{document}